%% file: paper-main.tex
\documentclass[a4paper]{article}
\usepackage{blindtext, graphicx, fullpage}
\usepackage{subfigure}

\usepackage[justification=centering]{caption}

\usepackage{filecontents}
\usepackage[noadjust]{cite}
\usepackage{multirow}
\usepackage{enumerate}
\usepackage{ucs,enumerate}
\usepackage{multirow}
\usepackage{booktabs}
\usepackage{color,soul}
\usepackage{caption}
\usepackage{url}

\usepackage[cmex10]{amsmath}
\newcommand{\gaia}{\textsf{GAIA}\xspace}
\newcommand{\ieeeZ}{IEEE 802.15.4\xspace}

\newcommand{\wifi}{WiFi\xspace}
\newcommand{\myparagraph}{\smallskip\noindent\textbf}

\begin{document}
%
\title{On Mining IoT Data for Evaluating the Operation of Public Educational Buildings}

\author{Na Zhu, Aris Anagnostopoulos and Ioannis Chatzigiannakis \footnote{Department of Computer, Control and Management Engineering, Sapienza University of Rome, 
Rome, Italy. Email: zhu.1706409@studenti.uniroma1.it, \{aris,ichatz\}@diag.uniroma1.it}}

\maketitle

\begin{abstract}
Public educational systems operate thousands of buildings with vastly different characteristics in terms of size, age, location, construction, thermal behavior and user communities. Their strategic planning and sustainable operation is an extremely complex and requires quantitative evidence on the performance of buildings such as the interaction of indoor-outdoor environment. Internet of Things (IoT) deployments can provide the necessary data to evaluate, redesign and eventually improve the organizational and managerial measures. In this work a data mining approach is presented to analyze the sensor data collected over a period of 2 years from an IoT infrastructure deployed over 18 school buildings spread in Greece, Italy and Sweden. The real-world evaluation indicates that data mining on sensor data can provide critical insights to building managers and custodial staff about ways to lower a building’s energy footprint through effectively managing building operations. 
\end{abstract}

\section{Introduction}
\label{sec:intro}
\input{paper-introduction}

\section{Related work}
\label{sec:sota}
\input{paper-related}

\section{IoT Infrastructure}
\label{sec:iot}
\input{paper-infrastructure}

\section{High-level Data Analysis}
\label{sec:iot}
\input{paper-data}

\section{Thermal Comfort of Classrooms}
\label{sec:comfort}
\input{paper-comfort}

\section{Classroom Thermal Performance}
\label{sec:temperature}
\input{paper-temperature}

\section{Conclusions}
\label{sec:conlusion}
\input{paper-conclusions}

\section{Acknowledgments}

This work has been partially supported by the EU research project ``Green Awareness In Action'' (GAIA), funded under contract number 696029 and the research project Designing Human-Agent Collectives for Sustainable Future Societies (C26A15TXCF) of Sapienza University of Rome. This document reflects only the authors' view and the EC and EASME are not responsible for any use that may be made of the information it contains. 

A previous version of this paper has appeared in the \textit{2018 IEEE International Conference on Pervasive Computing and Communications Workshops (PERCOM Workshops)}, DOI: \url{https://doi.org/10.1109/PERCOMW.2018.8480226}, \cite{8480226}. \textcopyright 2018 IEEE.  Personal use of this material is permitted.  Permission from IEEE must be obtained for all other uses, in any current or future media, including reprinting/republishing this material for advertising or promotional purposes, creating new collective works, for resale or redistribution to servers or lists, or reuse of any copyrighted component of this work in other works.


\end{document}

%% file: paper-introduction.tex
Public educational systems on a national level involve the operation and management of a massive number of buildings that possess vastly different characteristics (size, age, location, construction, thermal behavior). The energy and environmental impact caused by public educational buildings via their complex activities and operations in teaching and research, as well as provision of support services could be considerably reduced by an effective choice of organizational and managerial measures. An interesting aspect regarding the energy efficiency of schools is the fact that historically, energy expenses in public educational systems have been treated as relatively fixed and inevitable. Evidence shows that a focus on energy use in schools yields an array of important rewards in concert with educational excellence and a healthful learning environment~\cite{powerdown}. Since energy costs are the second largest expenditure within public educational systems budgets, exceeded only by personnel costs~\cite{epa}, significant savings can be carved out, if energy consumption can be reduced. In particular space heating and cooling accounts for nearly 20\% of all energy consumption in the US~\cite{usde}. An important factor in the heating and cooling of classrooms is the orientation of the buildings. An optimal orientation creates opportunities to utilize the potential contributions of the sun, topography, and existing vegetation for increased energy efficiency by maximizing heat gain (or minimizing heat loss) in winter and minimizing heat gain in summer. In the case of existing buildings, arrangement of interior spaces, strategic landscaping, and modifications to the building envelope can mitigate unfavorable orientation. 

Although numerous studies have been conducted in terms of the arrangement of interior and outdoor spaces to achieve optimal natural lighting and maximize the effect of sun on heating, there are no quantitative evidence on the performance of school buildings using real-world measurements. Without evidence on the performance of buildings such as the interaction of indoor-outdoor environment, strategic planning and sustainable operation of educational buildings becomes extremely difficult. It is therefore critical to provide real-world measurements on the actual conditions of classrooms and in general school buildings in terms of the ambient conditions. In this sense, deploying an IoT infrastructure will provide the necessary data to evaluate, redesign and eventually improve the organizational and managerial measures. Employing data mining techniques over sensor data will provide critical insights to building managers and custodial staff about ways to lower a building’s energy footprint through effectively managing building operations~\cite{cross12,schelly11,schelly12}

In this work the data collected from the \gaia platform~\cite{s17102296,mylonas2018enabling,amaxilatis2017enabling,akrivopoulos2018fog} are analyzed in order to analyse the indoor conditions of school buildings across Europe. The \gaia platform has deployed a pilot IoT infrastructure spread in 3 countries (Greece, Italy, Sweden), monitoring in real-time 18 school buildings in terms of electricity consumption and indoor and outdoor environmental conditions. The data collected over a period of 2 years is analyzed in order to determine the indoor conditions of classrooms and provide insights on the operation of these educational buildings. This is a first attempt to establish a comparative evaluation of different buildings and classrooms using quantitative measures. The analysis presented here can help building managers to quickly identify classrooms that do not have optimal indoor environmental conditions and proceed by conducting targeted interventions to improve the conditions. It is an important tool that can assist in reducing the energy consumption in the long-term and in this sense contribute towards the sustainability of public educational buildings.

The rest of the paper is structured as follows. In Sec.~\ref{sec:sota} relevant literature is presented and in Sec.~\ref{sec:iot} the IoT infrastructure deployed by \gaia platform is presented in details. In Sec.~\ref{sec:iot} a high-level analysis of the data is presented and then in Sec.~\ref{sec:comfort} the thermal comfort of classrooms is evaluated in a way such that it can be comparatively evaluated with other sites of similar characteristics. In Sec.~\ref{sec:temperature} the indoor environmental conditions of classrooms is analyzed in terms of thermal conditions and an automated method is presented to identify classrooms with poor thermal performance. In The paper concludes in Sec.~\ref{sec:conlusion} where future research directions are also provided.

%% file: paper-related.tex
When it comes to private residential energy monitoring solutions, several prototypes for domestic power consumption estimation have been presented \cite{karthikeyan2017}. A more holistic approach to real-time building energy modeling through IoT device integration along with structural information extraction per building from dedicated databases is presented in \cite{bottaccioli2017}. On the other hand, the analysis the performance of non-residential buildings is yet an unexplored area of research. The work of \cite{MILLER2017360} examines the energy consumption of university campuses from around the world and provides a series of data mining techniques to reduce the expert intervention needed to utilize measured raw data in order to infer information such as building use type, performance class, and operational behavior. The work presented here follows a similar path, however focusing on the environmental conditions of classrooms and school buildings rather than the energy consumption.

More recently, \cite{ALLAB2017202} studied for the first time the thermal comfort of a university buiding in relation to the energy consumption. The analysis aims to provide recommendations that i) reduce energy consumptions and ii) improve comfort level. This work follows a combined approach where simulation tools are used to predict the thermal comfort based on experiments conducted during specific periods of the year to evaluate parameters of the building envelope that contribute to the models used. In contrast to this approach, the work presented here uses data provided by an already deployed IoT infrastructure which has measures specific parameters of indoor environmental condition. In this sense the approach followed here aims to provide a generic methodology that can be reproduced to other school buildings at low cost and can thus scale to large number of buildings. 

The impact of air quality on concentration levels in classroom situation was studied in~\cite{BAKOBIRO2012215}. The concentrations of carbon dioxide and other parameters were monitored for three weeks in two selected classrooms in 8 primary schools in England. A selected set of interventions were made to improve the ventilation rate and maintain the temperature within an acceptable range using a purpose-built portable mechanical ventilation system. The investigation provided strong evidence that low ventilation rates in classrooms significantly reduce pupils’ attention and vigilance, and negatively affect memory and concentration. Our approach aims to provide a monitoring framework that allow the continuous evaluation of the conditions of classrooms.

%% file: paper-infrastructure.tex
\begin{figure}[!t]
\centering
\includegraphics[width=\columnwidth]{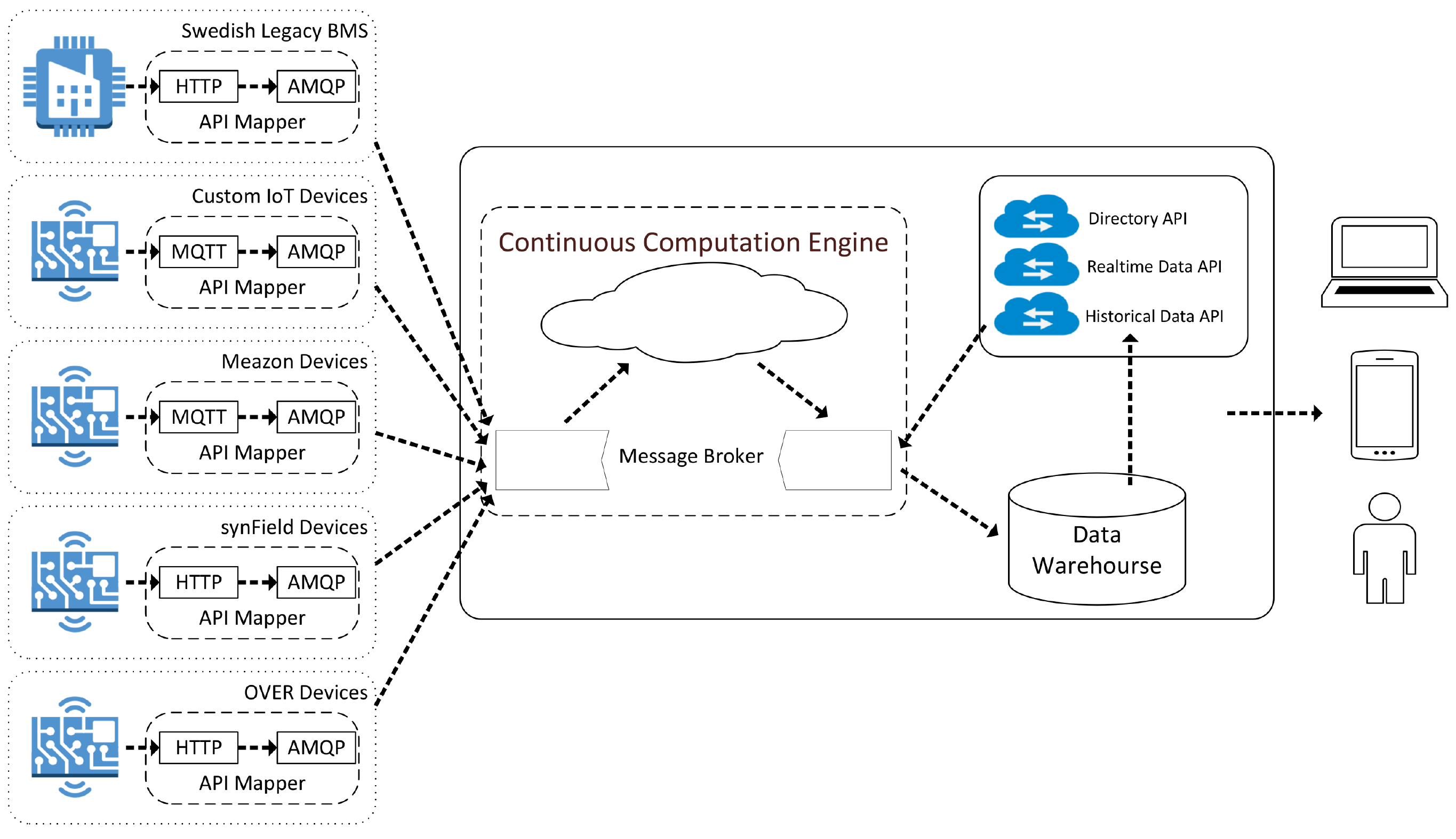}
\caption{\label{fig-gaia} Educational Building specific IoT architecture}
\end{figure}

The \gaia platform~\cite{s17102296} provides real-time monitoring of 18 school buildings (see Table \ref{table:table-schools}), spread in 3 countries (Greece, Italy, Sweden), covering a range of local climatic conditions and educational levels (primary, secondary and high school). In each of the 18 buildings participating in the \textsf{GAIA} platform, sensor devices are deployed that measure (a) the overall power consumption of the building, (b) the environmental comfort within each individual classroom and (c) the weather conditions and air pollution levels in each building. The diverse capabilities of IoT devices can be accessed via open interfaces for application developers. The platform's interfaces provide access to the available information in a way that suits all of the diverse end-users group that inherently exist in the educational sector: students, educators, building administrators and other administrative staff. For a graphical representation of the components used in the specific implementation fo the educational buildings see Fig.~\ref{fig-gaia}. For a detailed introduction to the platform, the provided services and the implemented applications see~\cite{s17102296,citeulike:12345348,Chatzigiannakis2005376,1612844,ewsn10}.

\begin{table}
    \centering
        \caption{Key facts of our deployment}
        \label{table:table-schools}
        \begin{tabular}{|l|c|l|}
          \hline
            \textbf{Parameter} & \textbf{\# } & \textbf{Description}\\
            \hline
            \multirow{2}{*}{Educational Buildings} & \multirow{2}{*}{18 } & 13 Greece, 4 Italy,\\
            &  & 1 Sweden\\
            \hline
            Sensing Points & 725 & $\ge$ 5 sensors per device\\
            \hline
            Students & 5500 & students in all levels\\
            \hline
            Teachers & 900 & teachers in all levels \\
            \hline
            Sensing Rate &  30 sec & classroom sensors \\
            \hline
        \end{tabular}
\end{table}

The IoT devices can be split into 3 different categories based on their origin and operation type:

\myparagraph{Power Consumption}
These devices are situated on the general electricity distribution board of each building to measure the power consumption of each one of the 3-phase power supply. They are networked via an \ieeeZ network and transmit the measurements to cloud services via \ieeeZ gateway nodes. In some location open-design devices are used (see~\cite{Pocero2017,Chatzigiannakis:2005:APC:1073970.1073985,Akribopoulos:2010:WSA:1908638.1909554}) and in some others off-the-shelf power meters are installed (e.g., produced by \textsf{Meazon}\footnote{http://meazon.com/}). 

\myparagraph{Weather and Atmosphere Stations}
These devices monitor atmospheric pressure and concentration of selected pollutants (to provide insights on the pollution levels) and measure weather conditions such as precipitation levels, wind speed, and direction. They are networked via Ethernet and powered using Power-Over-Ethernet, or communicate via \wifi and are either plugged into the sockets of the building or conduct energy harvesting via solar panels. Again, in some locations open-design devices are used (see~\cite{Pocero2017}) and in some others off-the-shelf weather stations are installed (e.g., \textsf{synField}\footnote{http://synfield.synelixis.com/}).  

\myparagraph{Environmental Comfort}
These devices measure various aspects affecting the well-being of the building's inhabitants, such as thermal (satisfaction with surrounding thermal conditions), visual (perception of available light) comfort and overall noise exposure. They also monitor room occupancy using passive infrared sensors (PIR). They are networked via an \ieeeZ network and transmit the measurements to the cloud services via \ieeeZ gateway nodes. Technical specifications for the devices can be found here~\cite{Pocero2017}.

%% file: paper-data.tex
\begin{figure}
\centering
\includegraphics[width=\columnwidth]{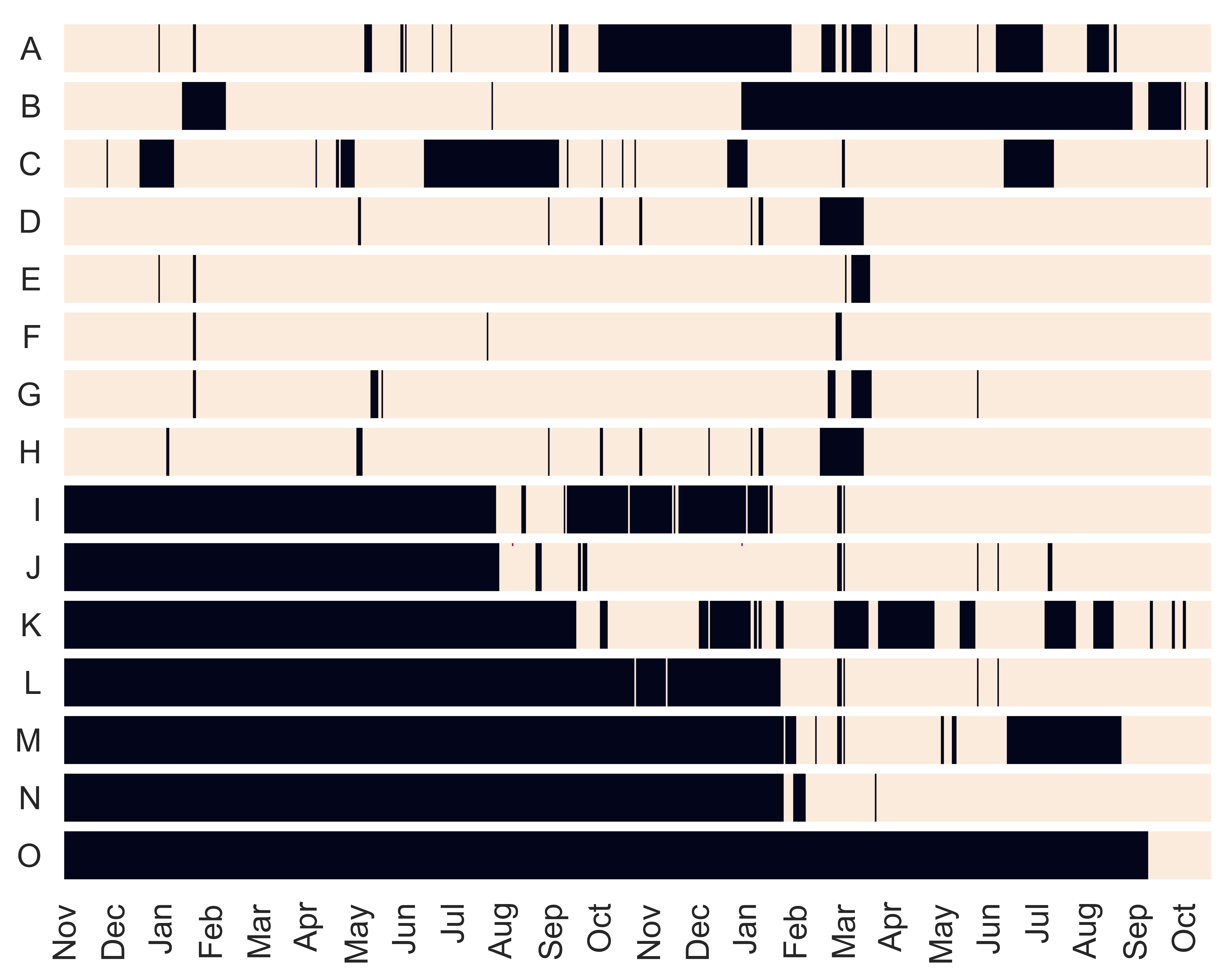}
\caption{\label{fig-data-1} Data availability per School Building}
\end{figure}

The IoT deployment of the \gaia platform is an ongoing process that
continuously expands to include additional school buildings. 
The collected data was analyzed to determine the variety of the
sensors supported and their points of sensing (POS), the velocity of
data arriving at the cloud infrastructure, as well as the variability of
the data collected. During this initial high-level analysis it became
apparent that the data collected from the IoT devices was not always
delived properly to the cloud. As a first step towards understanding the
availability of measurements, Fig.~\ref{fig-data-1} is included that
depicts the availability of measurements on a daily basis, organized based on the site
of deployment. A single dark point signifies missing data for the
specific on a specific day, while a white part signifies complete
availability (i.e., according to the sensing rate of the specific
sensor, see Tab.~\ref{table:table-schools}). This visualization shows the stability of
the IoT deployment. In almost all deployments there are values missing
almost on a daily level. Essentially these measurements are missing
either because they were never reported to the cloud infrastructure 
or due to a failure occuring while they were processed and stored by
the cloud services. In the first case, network failures occur either
due to a packet transmission error at the wireless network level (i.e., \ieeeZ  or \wifi), 
or due to a transmission error while an intermediate gateway transmitted the data
to the cloud infrastructure over the Internet. Since the access to the
measurements are done through the \gaia platform API, the reason for the
missing information is completely unknown. However, as it will become
evident in the following sections, specific data mining techniques can be
used to overcome the problem of missing values.

A second step for the high-level analysis is to examine the actual
values received from the IoT devices. It is very common in the relevant
literature to deploy relatively low-cost devices that produce
low-quality measurements or are not properly callibrated. For this
reason we examined the values to identify possible outliers, that is
observation points that are distant from the historic values. Such
observations may be due to transient errors occurring on the sensing
equipment and should be excluded from the data set. The identification
of outliers is based on the \textit{interquartile range} ($IQR$) using
the upper and lower quartiles $Q_3$ (75th percentile) and $Q_1$ (25th
percentile). The lower boundary is set to $Q_1 - 3 \times IQR$ and
the upper bounder is set to $Q_3 + 3 \times IQR$ where $IQR = Q_3 -
Q_1$.
The values are examined per sensor/site basis using a timed-window of
size $W$. If the a value is outside the boundaries $\left[Q_1 - 3 \times
IQR, Q_3 + 3 \times IQR\right]$ it is flagged as an outlier. In the
following sections we replace it with the minimum or the maximum value
observed during the time window $W$. After examining the values characterized
as outliers two distinct cases were identified: (a) $0$ values which
were clearly sensor error rather than natural events (e.g. humidity of $0\%$,
temperature dropping from $\sim 20$ to $0$), and (b) drastic changes
of power consumption (i.e., spikes or fast drops) that could not be justified
by the daily school activities.


In Table~\ref{table:table-data} the different school sites are
summarized indicating the time when they were incorporated in the \gaia
platform. For each school building the number of points of sensing (POS)
are listed along with the total number of sensors deployed. The table
reports the percentage of outages recorded for the particular site
(reflecting the periods during which no measurements were received
from the site, as a percentage from the point when the site was first
incorporated in the \gaia platform) along with the total number of 
measurements received from
this site is listed along with the percentage of values that have been
identified as outliers. Remark that to avoid issues related the
confidentiality of private data, the names of the school buildings have
been omitted. 

\begin{table}
\centering
\caption{Data availability per School Building}
\label{table:table-data}
\begin{tabular}{|c|c|c|c|r|c|}
\hline
Site & POS & Sensors & Start time & Outages & Outliers \\\hline
A  & 8 & 43 & 2015-OCT & 17.78\% & 2.67\% \\
B & 9 & 56 & 2015-OCT & 28.26\% & 2.84\% \\
C & 6 & 32 & 2015-OCT & 13.63\% & 2.82\% \\
D & 9 & 50 & 2015-OCT & 2.33\% & 1.45\% \\
E & 7 & 43 & 2015-OCT & 1.19\% & 1.63\% \\
F & 11 & 54 & 2015-OCT & 0.96\% & 3.36\% \\
G & 7 & 45 & 2015-OCT & 1.69\% & 1.33\% \\
H & 6 & 27 & 2015-OCT & 6.01\% & 1.31\% \\
I & 7 & 36 & 2016-SEP & 12.23\% & 1.68\% \\
J & 34 & 103 & 2016-APR & 3.97\% & 3.19\% \\
K & 5 & 26 & 2016-SEP & 15.87\% & 1.7\% \\
L & 5 & 109 & 2016-OCT & 36.43\% & 1.22\% \\
M & 5 & 24 & 2017-FEB & 23.09\% & 2.67\% \\
N & 12 & 55 & 2017-FEB & 1.31\% & 5.09\% \\
O & 4 & 22 & 2017-SEP & 0\% & 13.22\% \\ \hline
\end{tabular}
\end{table}

\begin{figure}
\centering
\includegraphics[width=\columnwidth]{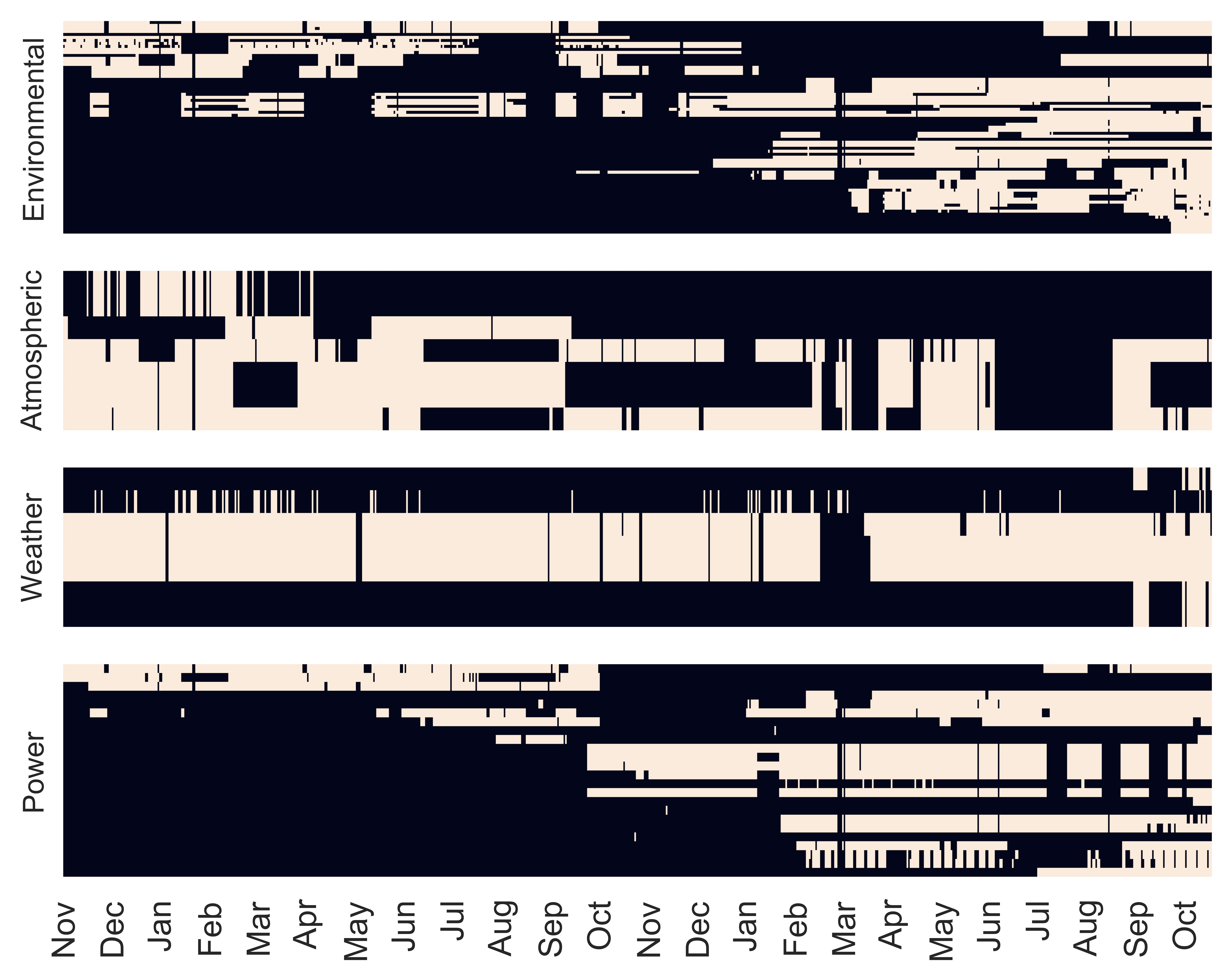}
\caption{\label{fig-data-2} Data availability per Sensor Type}
\end{figure}

The analysis reveals that certain buildings experience very
often data outages. At a second level the same analysis of data is
repeated based on the device type of the sensor. For each device category the percentage of outages and the percentage of
outliers observed are reported in Table~\ref{table:table-device}. 
In Fig.~\ref{fig-data-2} the representation of the availability of data
is depicted on a daily basis for each sensor separately organized by
sensor category. Based on this visualization one observes that all sensors
experience period loss of data. This may be justified by the wireless
networking technology used to interconnect the sensors located in the classrooms as
reported in~\cite{s17102296}. Apparently the low-power, lossy nature of 
the networking technologies used result to a non-negligible data lost.

\begin{table}
\centering
\caption{Data availability per Sensor Type}
\label{table:table-device}
\begin{tabular}{|l|c|c|r|r|}
\hline
Name 		& POS 	& Sensors 	& Inactive 	& Outlier \\ \hline
Environmental	& 101 	& 505 		& 14.62\%	& 7.76\% \\ 
Atmospheric  	& 7 	& 56 		& 19.56\% 	& 6.29\% \\
Weather 	& 7 	& 28		& 20.25\% 	& 0.95\% \\
Power  		& 20 	& 56 		& 12.55\% 	& 4.17\% \\ \hline
\end{tabular}
\end{table}

\begin{figure}[!b]
\centering
\includegraphics[width=.9\columnwidth]{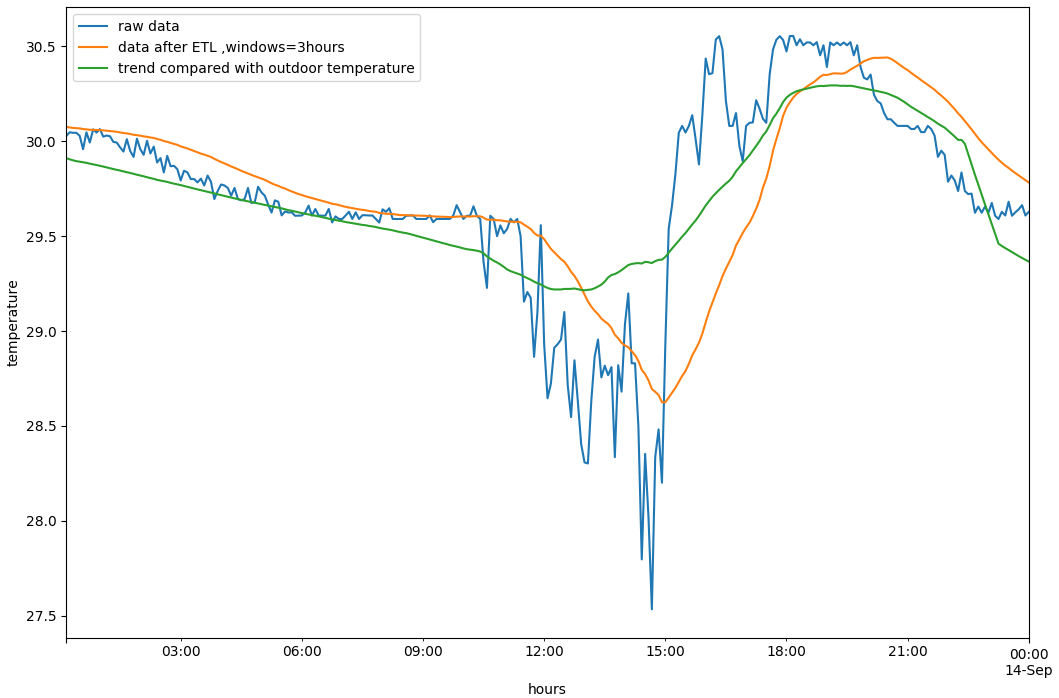}
\caption{\label{fig:etl} Temperature sensor time series processing}
\end{figure}

To overcome the fact that the IoT
deployment (1) experiences outages on a regular basis and (2) at a
significantly lower rate, sensors report values characterized as
outliers a moving window average technique is used. 
The moving window all to (1) smooths out short-term flactuations
for the case of outliers and (2) fills-in missing values using a simple
local algorithm that introduces historic values to fill in the missing
data for the specific time period. The moving window average also helps
to highlight longer-term trends on the sensor values.
Fig.~\ref{fig:etl} depicts an example of the analysis conducted over a
specific temperature sensor located in a classroom.

%% file: paper-comfort.tex
\begin{figure}[!t]
\centering
\includegraphics[width=\columnwidth]{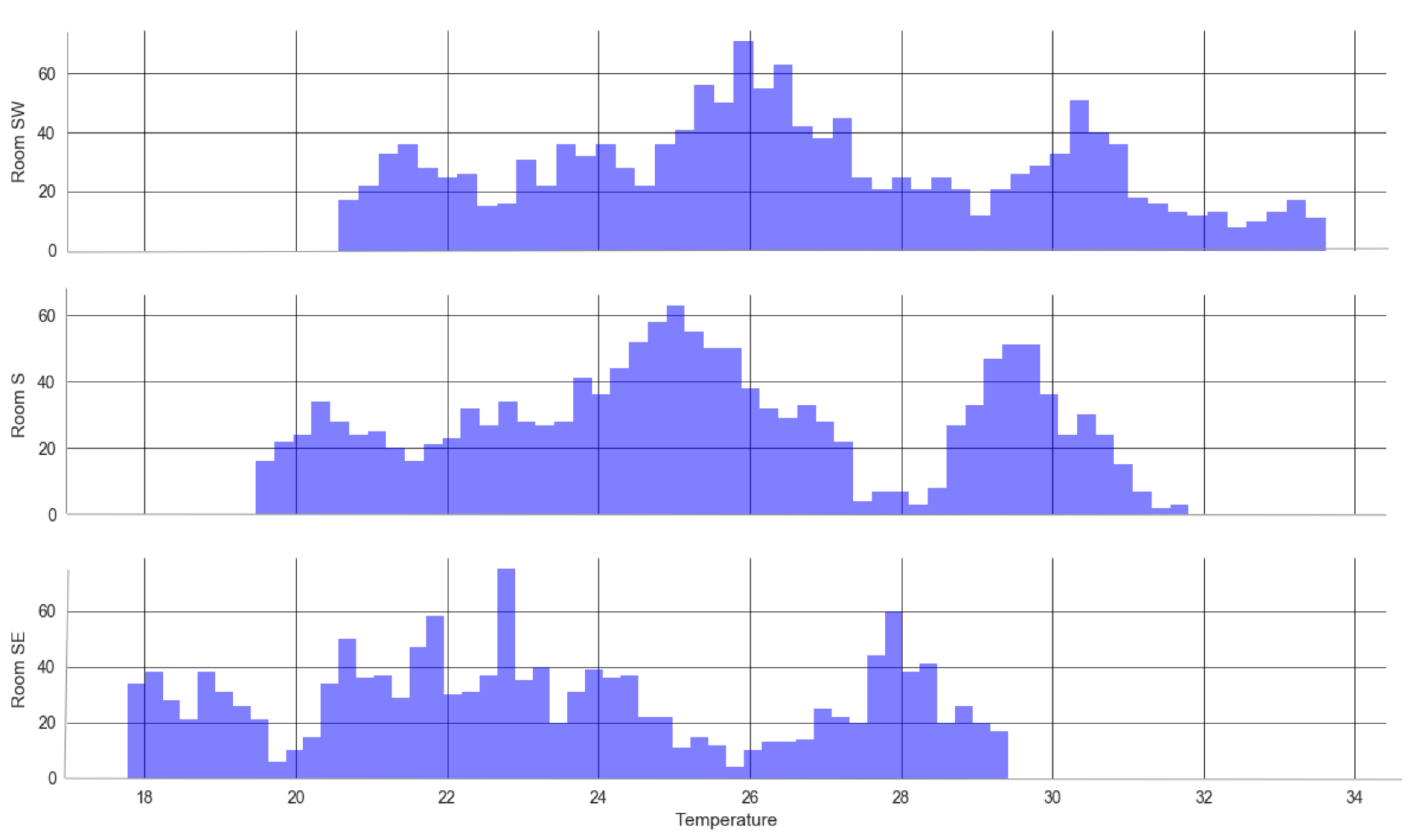}
\caption{\label{fig:histogram} Indoor temperature histogram for three classrooms during Sep/17 to Oct/17}
\end{figure}

Examining the classrooms temperature is a measure of understanding the
conditions under which students and teachers operate. Hot, stuffy
rooms---and cold, drafty ones---reduce attention span and limit
productivity. In Fig.~\ref{fig:histogram} a histogram is provided for the indoor
temperature of the three classrooms (facing south,
south-west and south-east) examined during a period of 2 months. 
Lower temperatures are observed in the room facing South-East in contrast to the other two rooms. 

Evaluating the indoor conditions requires considering also other factors 
related to the environment, such as humidity 
(e.g., excessively high humidity levels can also contribute to
mold and mildew), as well as what the students are wearing (which
depends on the period of the year). 

A common approach to examine the indoor conditions of the classrooms is
in terms of the \emph{thermal comfort}. The ANSI/ASHRAE
Standard 55, Thermal Environmental Conditions for Human Occupancy~\cite{ashrae} 
is defined to specify the
combinations of indoor thermal environmental factors and personal
factors that will produce thermal environmental conditions acceptable to
the majority of the occupants of an area. Thermal comfort is
primarily a function of the temperature and relative humidity in a room,
but many other factors affect it, such as the airspeed and the
temperature of the surrounding surfaces.
Furthermore, thermal comfort is strongly influenced by how a specific room
is designed (e.g., amount of heat the walls and roof gain or
lose, amount of sunlight the windows let in, whether the windows can
be opened or not) and the effectiveness of the HVAC system. 

The analysis presented here is done
based only on the quantitative information provided by the IoT
deployment of the \gaia project, we applied a simple approach that is
able to provide an estimate under the lack of all the necessary data.
For this reason the CBE Thermal Comfort Tool for ASHRAE-55\cite{cbe} 
was used to evaluate the thermal comfort of classrooms. 
Under the \emph{adaptive method} provided, thermal comfort is computed as a function of the indoor
temperature, the outdoor temperature along with the outside air wind
speed. The outside conditions (temperature, airspeed) were acquired by
the \emph{WeatherMap} service that also provides historical data. For
each different classroom, the thermal comfort is computed for each
different hour during the operation of the school (from 08:30 to 16:30). 
In the sequel, the individual values are averaged over each different day. 
A classroom with a daily comfort
of $1.0$ signifies that during all hours the conditions where within the
comfort zone defined by this particular formula, while a daily comfort
of $0.0$ signifies that during all hours the conditions were outside the
comfort zone. 

\begin{figure}[!t]
\centering
\includegraphics[width=\columnwidth]{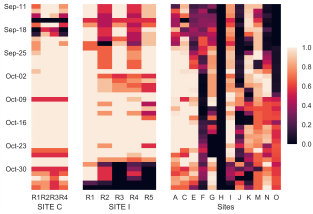}
\caption{\label{fig:comfort} Site thermal comfort during Sep/17 to Oct/17}
\end{figure}

On the right side of Fig.~\ref{fig:comfort} a summary of all the sites
participating in the \gaia platform is provided for a period of
two months (from Sep/2017 to Oct/2017). On the left side, the individual thermal
comfort of the classrooms of site $C$ is compared with those of site $I$.
Site $C$ achieved the highest comfort in contrast to site $I$ that achieved the
lowest. One reason for the difference is the actual location of the school,
site $C$ being the school located on the southern point in contrast to site $I$
which is one of the northern points. However, apart from the external weather conditions, there are other reasons that affect thermal comfort, e.g., such as the construction materials.
In the following section, data mining techniques are applied in order to identify in more
details classrooms with poor performance or user-related activities that may also affect thermal
comfort.

%% file: paper-temperature.tex

Several factors affect the internal classroom temperature
ranging from the local weather conditions to the orientation of the
room, the construction materials used (e.g., the insulation, windows)
and also the position of the radiators within the classrooms. The \gaia
platform includes school buildings located in different climatic zones,
constructed in different years ranging from 1950 to 2000, using diverse
materials and with different heating and ventilation systems.
Unfortunately, such information is not available in an open format so
that they can be incorporated via data integration techniques. In this
section, data mining techniques are used to derive information regarding
the thermal efficiency of the classrooms using the quantitative data collected from
the IoT deployment. 

The goal of the analysis is to identify classrooms with poor thermal performance. 
One of the leading factors affecting the temperature of the classrooms is their
orientation and the contributions of the sun towards the internal environment quality.
During a sunny day, it is observed that at mid-day the classrooms experience the highest temperatures.
It is also observed (see also Fig.~\ref{fig:histogram}) those classrooms facing south-west are exposed for longer periods of sun thus maintaining higher temperatures for longer
periods than classrooms with a different orientation. In order to factor in the potential contribution of the sun,
the time-series of the temperature sensors are examined in correlation with the external temperature, the cloud cover as well as the orientation of the classrooms.

A second important factor that affects the internal temperature of classrooms is the
daily activities of the students and teachers. Opening and closing the door, the windows and the window blinds have an immediate effect on the temperature. For example, when a window is open there
is a temperature drop of about $2^oC$. In order to overcome these effects, the performance of the
classrooms is examined only during weekends when there are no school activities.

\begin{figure}
\centering
\includegraphics[width=\columnwidth]{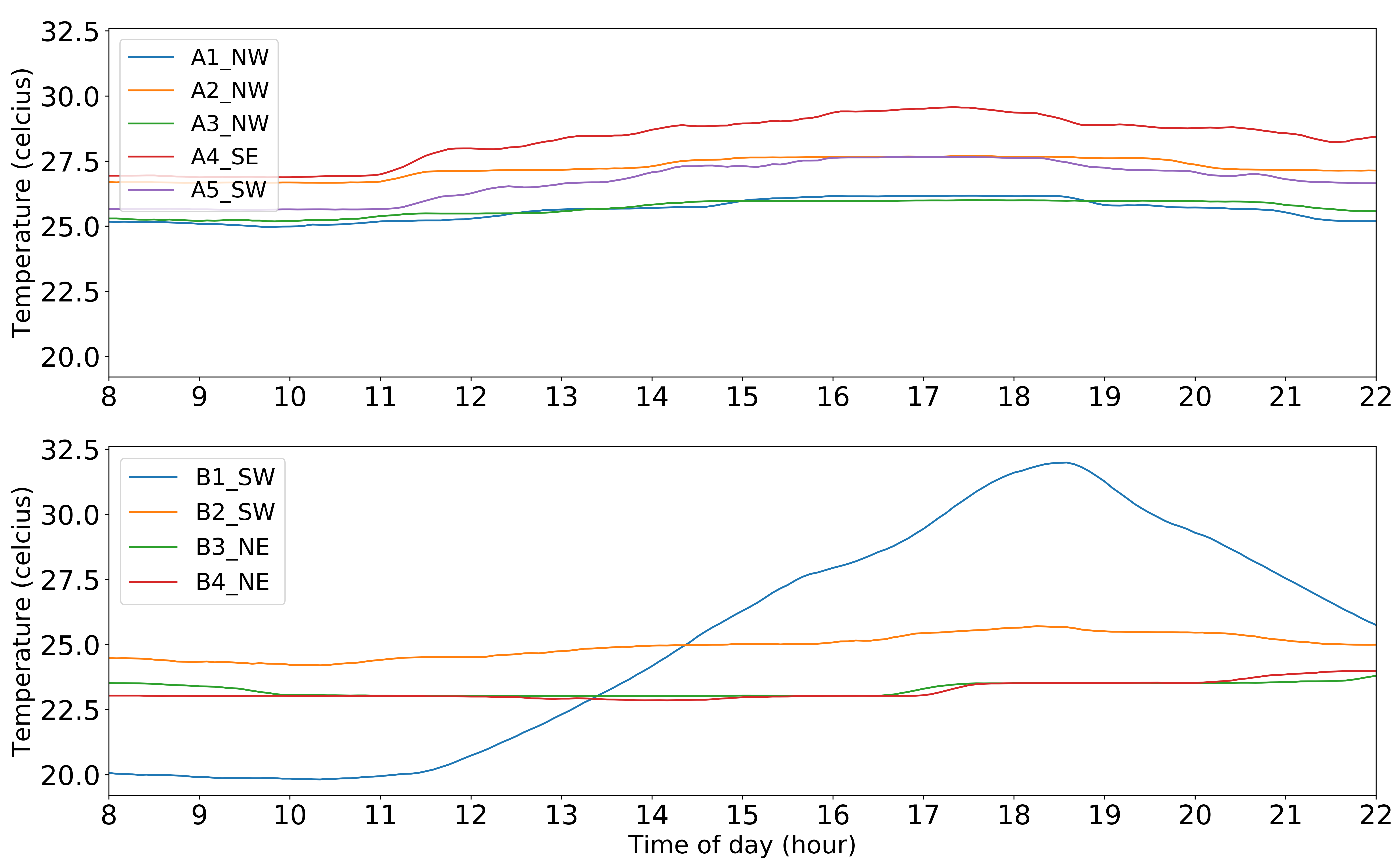}
\caption{\label{fig:weekend} Classroom temperature during 30/Sep (Saturday)}
\end{figure}

Given the above considerations, the temperature of each room is examined to identify
poorly performing classrooms. In Fig.~\ref{fig:weekend} two specific performance issues regarding two schools located in the same city are depicted. The first issue is related to the bottom figure, 
where room R1 achieves very poor performance with temperature starting at the very low level of $20^oC$ 
and increasing up to $32^oC$ within 8 hours. The second issue is related with the top figure, 
where the south-west facing classroom (R5) and the south-east facing classroom (R4) have an increase of 2 degrees during the day while all the other rooms are not affected. Even the south-west
facing room R2 of the bottom figure does not have such an increase during the day. 
After contacting the school building managers it was reported that (a) room R1 (bottom school)
is located outside the main building, within a prefabricated iso box where insulation is very poor and
(b) rooms of top school have no window blinds installed in contrast to the bottom school where window blinds
are installed in all rooms. These are just two examples of the results of the analysis conducted. 
It is expected that such an analysis can provide strong evidence on how to improve the performance of schools.

%
%
%
%
%
%

%% file: paper-conclusions.tex
In this work, a data mining approach has been presented that studies the indoor temperature of classrooms based on real-world data collected over a period of two years. Tools are provided to determine the thermal comfort of classrooms, identify classrooms with poor thermal performance and user activities that can affect the indoor quality of classrooms. This is the first quantitative analysis of the interaction of indoor-outdoor environment in educational systems. The results obtained provide critical insights to building managers about ways to improve classroom conditions by the arrangement of indoor spaces, strategic landscaping, and modifications to the building envelope. A natural next step would be to investigate the conditions in terms of lighting, air quality and noise thus providing a better understanding based on the Internal Environment Quality (IEQ) guidelines.